# Ultrafast carrier-coupled interlayer contraction, coherent intralayer motions, and phonon thermalization dynamics of black phosphorus

Mazhar Chebl, Xing He, Ding-Shyue Yang*

*Department of Chemistry, University of Houston, Houston, Texas, 77204, U.S.A.*

*Corresponding author. Email: yang@uh.edu

**Abstract**

Black phosphorus (BP) exhibits highly anisotropic properties and dynamical behavior that are unique even among two-dimensional and van der Waals (vdW) layered materials. Here, we show that an interlayer lattice contraction and concerted, symmetric intralayer vibrations take place concurrently within few picoseconds following the photoinjection and relaxation of carriers, using ultrafast electron diffraction in the reflection geometry to probe the out-of-plane motions. A strong coupling between the photocarriers and BP's puckered structure, with the alignment of the electronic band structure, is at work for such directional atomic motions without a photoinduced phase transition. Three temporal regimes can be identified for the phonon thermalization dynamics where a quasi-equilibrium without anisotropy is reached in about 50 ps, followed by propagation of coherent acoustic phonons and heat diffusion into the bulk. The early-time out-of-plane dynamics reported here have important implications for single- and few-layer BP and other vdW materials with strong electronic–lattice correlations.

**Introduction**

Besides the spanning of the entire spectrum of electronic properties from gapless graphene to insulating hexagonal boron nitride, two-dimensional (2D) materials and van der Waals (vdW) layered heterostructures have been found to exhibit exceptional quantum and correlated phenomena including superconductivity (*1*), fractal quantum Hall effects (*2, 3*), emergent ferromagnetism (*4*), and valley polarization (*5*). Understanding the interplay among the electronic, structural, and spin degrees of freedom and finding additional modulation and control by external stimuli have thus become prominent research endeavors and the foundations to explore the technological potentials of these materials. For black phosphorus (BP), significant attention has been on its highly anisotropic transport, optical, thermal, and mechanical properties (*6*) as well as the strong dependence of its band structure and correlated behavior on hydrostatic pressure (*7*), the stacking order (*8*), and the layer number (*9*). The puckered layers and structural anisotropy of BP in ambient conditions show prominent differences than planar graphene, as a result of every P atom having one more valence electron compared to carbon and forming a covalent bond with three neighboring atoms (Fig. 1A). Electronically, unlike many transition metal dichalcogenides (TMDs), the direct bandgap is maintained for all layer numbers of BP, which is at the Γ point for single to few layers and at the high-symmetry Z valley along the vdW cross-plane direction for the bulk (*10*). Such an evolution of the band structure is quite unique, which coincides with the finding of bond-like wavefunction overlaps between BP layers and their important role in an unusually strong interlayer coupling compared to the counterparts in graphene and TMDs (*11, 12*). Thus, BP as a layered material shows a profound connection between its structure and various properties in both covalently-bonded in-plane and vdW-separated cross-plane directions.

Dynamically, photoinduced responses of BP have been investigated by various time-resolved measurements, mostly concentrated on the in-plane directions. At low photoinjection densities, exciton–exciton annihilation governed the relaxation of carrier dynamics (*13*). Increased photocarriers cause bandgap renormalization as a result of many-body interactions (*14, 15*) as well as band filling and radiative recombination (*16*). Interestingly, a resembling carrier–phonon scattering time was found independent of the photoexcitation and probe polarizations, due to fast randomization of the carriers' distribution in *k*-space by carrier scattering (*17*). Moreover, the coherent phonon modulation observed in optical data has been associated with a breathing mode for few layers of BP and with longitudinal acoustic phonons for thicker specimens (*18, 19*). However, direct observations of the photoinduced structural dynamics require techniques such as ultrafast transmission electron microscopy (*20*) and diffraction (*21, 22*). The time-dependent changes of in-plane Bragg diffractions suggest a two-stage relaxation mechanism for fast electron–phonon coupling and slower thermalization of phonons; analysis of the in-plane diffuse electron scattering data reveals stronger carrier–phonon coupling along the zigzag direction at early times. Nonetheless, photoinduced structural dynamics and relaxation pathways in the vdW cross-plane direction remain unaddressed, which leaves out the important role of the interlayer coupling in BP in dynamics.

In this report, ultrafast electron diffraction (UED) in the reflection geometry is used to directly reveal the out-of-plane atomic motions of bulk BP. We show that an above-gap excitation triggers an initial interlayer contraction with a magnitude independent of the optical fluence used, whose lattice change has a strong carrier-coupled origin. More surprisingly, the onset of the diffraction intensity change is delayed by ~3 ps, which has not been observed for Bragg spots of materials with repeated intra-cell lattice changes but no involvement of a phase transition. It is found that such an onset delay can be attributed to the coherent motions of BP layers that match with the $A_g$ optical phonons. In the temporal range studied, three temporal regimes may then be identified to describe the carrier–phonon scattering and phonon thermalization processes.

**Results**

Photoexcitation of BP is achieved by using 2.41-eV photons, where photocarriers are initially injected without particular preferences in the Brillouin zone (Fig. S1). At early times, carrier scatterings and energy relaxation toward the bandgap in the Z valley lead to the transfer of a significant amount of excess energy to the BP lattice, whose structural dynamics result in time-dependent diffraction changes (Fig. 1, B and C). The absence of noticeable intensity change near the shadow edge region indicates the negligible role of interference by surface transient electric field effects, if any. By varying the laser polarization along the in-plane directions (i.e., from the armchair *c* axis to the zigzag *a* axis), the magnitude of the diffraction intensity difference changes, whose dependence coincides well with that of the photoabsorption coefficient (Fig. 1D). Hereafter, the polarization is kept along the armchair direction in this work.

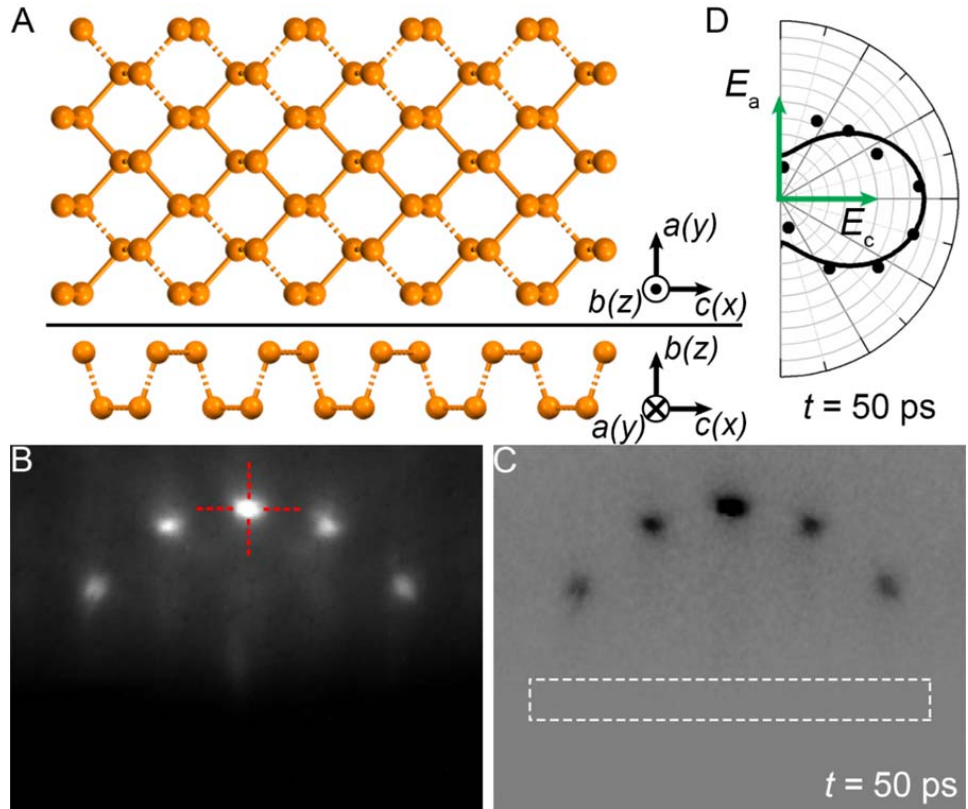

**Fig. 1. Structure and photoinduced diffraction changes of BP.** (**A**) Structure of BP in top (upper) and side (lower) views with the indicated crystal axes. (**B**) Diffraction image prior to photoexcitation. The profiles obtained along the vertical and horizontal dashed lines are fitted to a Lorentzian function to extract the intensity and position of the central spot. (**C**) Diffraction difference image at 50 ps referenced to the negative-time frame. No noticeable time-dependent changes are found in the shadow edge region indicated by the dashed box. (**D**) Polar plot of the intensity change observed at 50 ps as a function of the polarization varied in the horizontal *a*–*c* plane.

**Initial interlayer lattice contraction**

Instead of the Debye–Waller type of randomized atomic motions typically observed in photoexcited materials without a photoinduced phase transition, we find that BP exhibits unique directional motions along the vdW-stacked cross-plane direction within the duration of a few picoseconds (ps). Shown in Fig. 2A is the comparison of the diffraction intensity (*I*) and out-of-plane position (i.e., momentum transfer $s_\perp$) changes of the (0 16 0) spot. Two prominent features are worth noting. First, following photoexcitation, an initial positive change $\Delta s_\perp/s_\perp$ of $\sim 8\times 10^{-4}$ is observed at about 3 ps and precedes the reversed, long-term negative change at later times.

Such an observation indicates a lattice contraction $\Delta b/b$ prior to the dominance of the anticipated thermal expansion after 6 ps, which is reminiscent of the comparable behavior found in layered graphite (*23, 24*) and TMDs (*25*). However, the initial contraction $\Delta b$ appears to be independent of the photoinjection levels used here, whereas the later interlayer lattice expansion is proportional to the laser fluence (Fig. 2B).

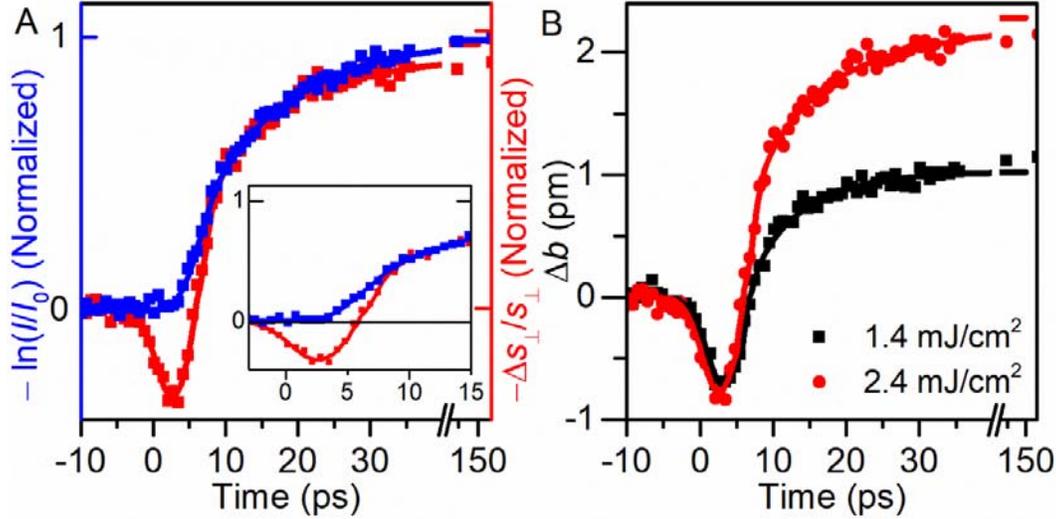

**Fig. 2. Photoinduced structural changes of BP.** (**A**) Normalized changes in the intensity and vertical position of the (0 16 0) spot. The solid lines are guides to the eye. The inset shows the early-time dynamics. (**B**) Corresponding lattice changes in the out-of-plane *b* axis at two different laser fluences used.

We note that the origin of BP's early-time interlayer contraction has distinct differences from those for graphite and TMDs, with a significant role played by the electronic structure of BP. It is known that BP in equilibrium has strong interlayer coupling due to the lone electron pairs pointing along the cross-plane direction, which results in a prominent wavefunction overlap beyond van der Waals interactions and makes the Z-valley bandgap quite unique among layered materials (*9, 11*). Dynamically, after their energy relaxation within one to few ps, the wavefunctions of near-gap photocarriers in the Z-valley have periodic modulations across the vdW-stacked layers. Consequently, the interlayer coupling is expected to be affected due to the presence of these photocarriers. With the out-of-plane elastic constant $C_{33} = 52.66$ GPa (*26, 27*) and the ellipsoidal volume of the anisotropic exciton extensions $V_{ex} \cong 120$ nm$^3$ (*28*), we find that the elastic energy induced in the contracted lattice is $\frac{1}{2}C_{33}(\Delta b/b)^2 V_{ex} \cong 10$ meV, which is comparable to the binding energy of free excitons in BP, about 8 to 9 meV (*28, 29*). Such an energy coincidence may be rationalized in the simple framework of a carrier-coupled displaced oscillator, hence implying an important carrier–lattice correlation. Moreover, the bandgap renormalization due to photocarriers' many-body effects can reach tens of meV in the fluence range used (*30*). These are in contrast with the limited electronic change predicted in absence of a strong photocarrier–lattice coupling. For example, theoretical results of unexcited BP with an interlayer distance change indicate, proportionally, a decrease of ≤ 2 meV in the bandgap for the observed lattice contraction (*31*); pure pressure-induced band renormalization via deformation potentials is also insufficient in understanding photoinduced responses of BP (see below).

**Early-time coherent intralayer vibrations**

The second prominent feature of the initial lattice dynamics is a delay of ~3 ps for the onset of the diffraction intensity change, which is entirely unexpected and also significant in the duration, compared with the early-time dynamics of graphite (*23*). Certainly, P atoms should acquire additional vertical movements as a result of sub-ps carrier–phonon coupling; a decrease in the diffraction intensity would have been the anticipated observation for BP without a photoinduced phase transition, as a result of the incoherent Debye–Waller randomized atomic motions (*21, 23, 32*). Thus, the lack of an intensity change requires the consideration of coherent intra-cell atomic motions at early times. Given the same element in BP, the intensity of the (0$k$0) spot is proportional to the square of the structure factor $F$ for a unit cell (omitting the contribution of thermal motions as a Debye–Waller factor at finite temperatures),

$$F = \sum_{j=1}^{8} f_j \, e^{-2\pi i k(z_j + \delta_j)} = f_P \sum_{j=1}^{8} e^{-2\pi i k(z_j + \delta_j)} \tag{1}$$

where $f_P$ is the atomic scattering factor of phosphorus and $z_j$ and $\delta_j$ are the out-of-plane fractional position and induced displacement of atom $j$ in a cell, respectively (Fig. 3A). Before photoexcitation with $\delta_j = 0$, the structure factor is $F_{eq} = f_P \cdot 8 \cos(2\pi k z)$ based on the puckered-layered structure of BP where $z_1 = z_2 = +z$, $z_3 = z_4 = -z$, $z_5 = z_6 = 1/2 + z$, and $z_7 = z_8 = 1/2 - z$ with $z = 0.10168$. After photoexcitation, concerted atomic motions following a symmetric $A_g$ phonon mode (out-of-plane $A_g^1$ and/or in-plane $A_g^2$ with small out-of-plane components) will lead to a modified structure factor $F = f_P \cdot 8 \cos[2\pi k(z + \delta)]$. For all other motions including the $B_{3g}^2$, $B_{1u}$, asymmetric (across adjacent layers) $A_g$, and breathing modes, the modified structure factor is $F = f_P \cdot 8 \cos(2\pi k z) \cos(2\pi k \delta)$.

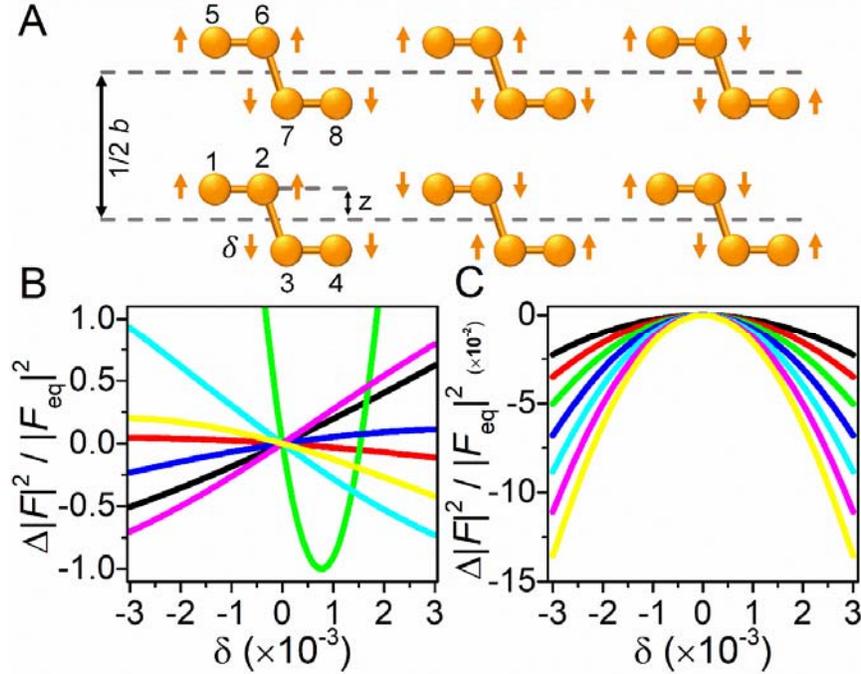

**Fig. 3. Coherent out-of-plane atomic movements of BP and expected diffraction intensity changes.** (**A**) Schematics of three intra-cell modes (with only the out-of-plane components indicated), which correspond to symmetric $A_g$, asymmetric $A_g$, and $B_{1u}$ (respectively from left to right) as examples. The orange arrows indicate the relative signs of the displacements in the $b$ (i.e., $z$) axis. (**B**) Intensity changes of the ($0k0$) diffraction spots as a function of the displacement amplitude corresponding to the symmetric $A_g$ motions (A, left). Different colors are used for even orders of $k$ from 8 to 20. (**C**) Intensity changes of the ($0k0$) spots for all the other intra-cell modes.

Shown in Fig. 3, B and C are the resulting diffraction intensity changes with a small $\delta$, where only a symmetric $A_g$ phonon mode gives essentially no intensity change (due to the quasi-linear relation between $\Delta|F|^2 / |F_{eq}|^2$ and $\delta$, averaged over an oscillation period of sub-100 fs faster than our temporal resolution) and the rest should result in an intensity decrease (more pronounced for higher diffraction orders). Thus, our observation is consistent with photocarriers' early-time preferential coupling to symmetric $A_g$ phonon modes. Together with the aforementioned compressive strain, the out-of-plane directional, concerted atomic motions add further evidence for the significant coupling of the Z-valley photocarriers to the out-of-plane lattice structure. We further note that the present results are also consistent with the in-plane observations for highly anisotropic transient phonon populations in the first few ps (*22*). In such a nonthermalized state, the structure factor analysis for ($h$00) and (00$l$) spots show that all intra-cell normal modes yield an intensity dependence the same as shown in Fig. 3C except for only the symmetric $A_g^2$ mode along the armchair direction for (00$l$) (Fig. S2). Consequently, the initial population of various in-plane phonons, especially along the zigzag direction, results in an early-time diffraction intensity decrease without a clear onset delay, and hence the conventional Debye–Waller model was still used although phonon thermalization is not reached (*21*).

**Three regimes of phonon thermalization**

Following the onset delay, the intensity of the (0 16 0) spot decreases to a minimum level corresponding to the fluence used within 50 ps (Fig. 4A). A biexponential function is more adequate to describe the dynamics compared to the fit of a single exponential function that yields more residual errors (Fig. 4B); a fast time constant of 0.9 to 1.0 ps and a slower one of 16±5 ps are obtained, independent of the photoinjected carrier density (the range and the standard deviation are assessed using the early-time data from 3 samples with 11 different laser fluences in total). These characteristic times closely resemble those reported for the in-plane structural dynamics (*21, 22*). Hence, the following picture with three major temporal ranges is reached. In the carrier-coupled regime for the first 3 ps, the relaxation of energetic photocarriers preferentially produces anisotropic excitation of phonons with momenta along the zigzag direction and the symmetric $A_g$ modes across the vdW-stacked layers. Concurrently, the resulting excitons and free carriers in the Z valley modulate the cross-plane electronic structure and induce an interlayer contraction as a result of strong photocarrier–lattice coupling. In the second regime up to 50 ps, thermalization of the nonequilibrium phonons becomes a dominant theme of the dynamics, with a transition time of 4–10 ps for still separated intensity and lattice expansion curves and an equilibration period after 10 ps for largely matched dynamics (Fig. 2A). Coherent acoustic phonons are also produced during this period, whose propagation into the bulk can be monitored by optical transient reflectivity (Fig. S3).

In the final regime after 50 ps, the lattice reached a quasi-equilibrium of thermalized phonons where an effective increase in the lattice temperature $\Delta T$ can be determined and the Debye–Waller model becomes suitable for the randomized atomic motions. We note that although the photocarriers have a lifetime covering all three regimes, the approximate energy they retain across the bandgap of 0.3 eV is a minor fraction compared to the much larger excess part (2.41 eV minus 0.3 eV per electron−hole pair) transferred earlier to the BP structure. Using the data at 50 ps obtained from 13 BP samples and a linear fit of $\Delta b/b = -\Delta s_\perp/s_\perp$ at different fluences, we obtain an out-of-plane lattice expansion coefficient of $\sim(5.0\pm0.4)\times10^{-5}$ K$^{-1}$ (Fig. 5A), which agrees reasonably with the literature value without the influence of photoinjected carriers (*33*). In addition, the relative changes in the diffraction intensity at different fluences $\ln[I(50\text{ ps})/I(t < 0)]/\Delta T$ is approximately $-3h^2 s_\perp^2/mk_B\Theta_D^2$ according to the Debye–Waller model above BP's Debye temperature $\Theta_D$, where $h$ is the Planck constant, $m$ the mass of a phosphorus atom, and $k_B$ the Boltzmann constant (see Supplementary Materials and Fig. S4). A linear dependence on $s_\perp^2$ is found in Fig. 5B and a fit gives an estimate of $\Theta_D^b \cong 250\pm13$ K. Furthermore, a cross-plane thermal conductivity of $\sim$2 W m$^{-1}$ K$^{-1}$ is obtained from a fit of the diffraction intensity changes on the sub-nanosecond scale considering one-dimensional thermal diffusion (see Supplementary Materials and Fig. S5).

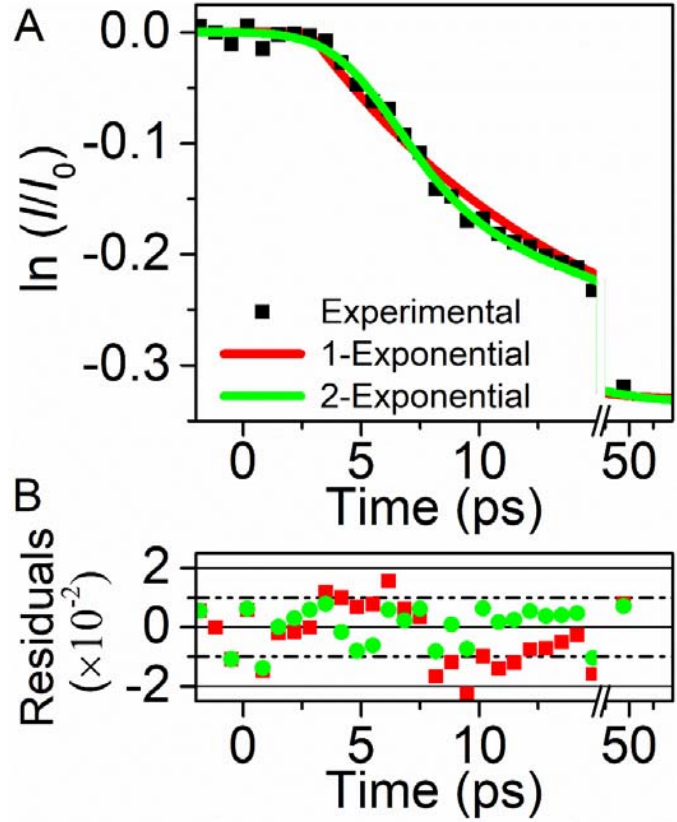

**Fig. 4. Early-time diffraction intensity change.** (**A**) Comparison of the fits with single- (red) and bi-exponential (green) functions. (**B**) Residuals from the two fit models. Larger systematic deviations are seen from the single-exponential fit.

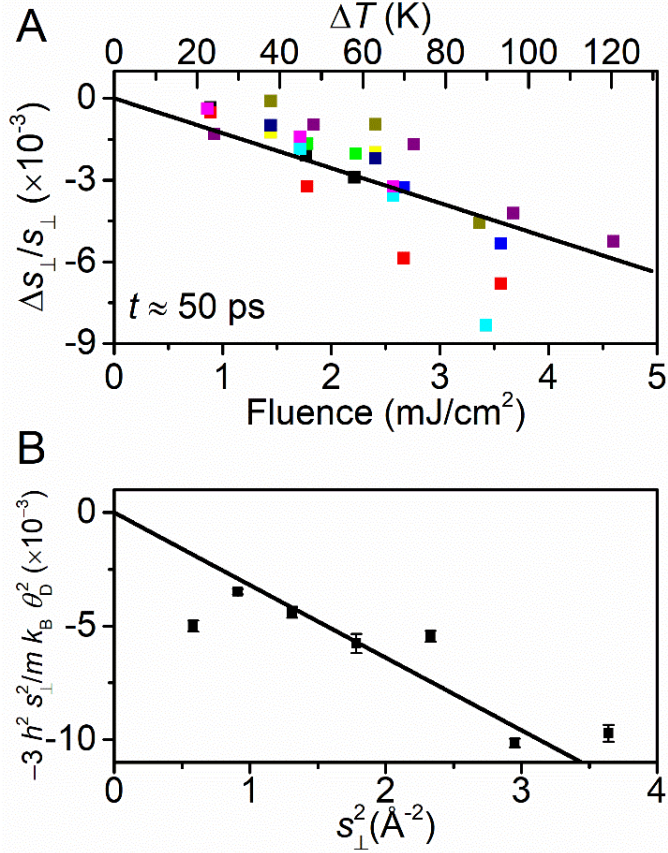

**Fig. 5. Long-time structural changes and phonon thermalization.** (**A**) Change of the diffraction spot position along the out-of-plane direction at ~50 ps as function of the laser fluence used and the estimated temperature jump. Results from the same sample are denoted in one color. The solid line is an overall linear fit. (**B**) Debye−Waller analysis of the diffraction intensity changes. Each data point corresponds to the slope extracted from the fluence-dependent data for the corresponding diffraction order.

**Discussion**

In addition to the exciton-coupled elastic energy estimated earlier, a compressive lattice strain of $8\times10^{-4}$ near the BP surface is equivalent to an applied pressure of $C_{33}(\Delta b/b) \cong 40$ MPa along the surface normal direction. For the comparable observations of TMDs, Mannebach *et al.* considered an analytical model to explain the dynamic modulation of the interlayer van der Waals interaction (*25*). According to the Lifshitz model with a Drude-like electron gas in two charged slabs, an interlayer contraction of $8\times10^{-4}$ generates an effective pressure of ~50 MPa at a near-surface carrier density of $2\times10^{20}$ cm$^{-3}$ in BP; a superlattice model considering the correlations between quantum fluctuations of optically excited electrons and holes yields a low value of ~7 MPa at the same carrier density (*25*). All these values are at least 1.5 orders of magnitude lower than the pressure required for the semiconductor-to-semimetal electronic transition (*34*), which validates that the observed photoinduced dynamics are perturbative in nature and far from the scenario with large-scale band structure changes. However, we note that the estimates based on the Lifshitz and superlattice models represent the upper bound. While the agreement of the photoinduced pressure suggests the usefulness of the picture given by these models, we caution that the high carrier density may see a significant reduction at 3 ps due to various decay channels (defect-related trapping, bimolecular and/or Auger recombination, exciton–exciton annihilation, etc.) as the photocarriers relax toward the narrow Z valley and therefore are not maintained at the initial injection level (*30, 35*). With a reduced carrier density (whose value may be highly dependent on the sample quality), the estimated pressure based on a Lifshitz-like model may be much lower and not fully consistent with the experimentally observed compressive strain. Taking the interlayer contraction and intralayer coherent atomic motions together, we consider the important role of BP's band alignment for cross-plane photocarrier–lattice coupling that does not exist in typical TMDs.

The current results elucidate early-time photocarrier-driven interlayer lattice contraction and out-of-plane vibrational motions within each puckered layer in a concerted, symmetric fashion, as well as the three major temporal regimes from directional to thermalized atomic motions for the photoinduced dynamics of BP. Its electronic structure with the bandgap at the Z valley aligning with the out-of-plane axis plays a critical role in the strong photocarrier–lattice coupling, which illustrates the significance of probing structural dynamics in the vdW-stacked direction. It will be essential to further examine BP with a reduced thickness to even a monolayer to reveal the impact of the electronic band structure change. As vdW-stacked heterostructures open up new opportunities with engineered phenomena and control, attention for the interlayer dynamic behavior is eminently needed especially when an alignment between the electronic structure and the lattice exists.

**Materials and Methods**

Crystalline BP samples were obtained from two sources, one from the University of Science and Technology of China (USTC) and the other from HQ graphene commercially. With mechanical exfoliation, the USTC crystals exhibit larger domains whereas the commercial samples produce flakes more of a needle-like shape. However, no major differences were found in their photoinduced dynamics. For each sample, a fresh surface was prepared by exfoliation of the bulk in air and quickly loaded into the ultrahigh vacuum (UHV) assembly with a base pressure of $2\times10^{-10}$ Torr in the main chamber to minimize surface deterioration over time.

Details about the reflection UED apparatus has been described previously (*32, 36*). In short, the 515 nm (2.41 eV) photoexcitation pulses were produced by second harmonic generation (SHG) of the fundamental output (1030 nm, 170 fs) of a Yb:KGW regenerative amplifier laser system. Another stage of SHG using a fraction of the 515-nm beam produced the ultraviolet (257 nm) pulses, which were then guided toward and focused on a $LaB_6$ emitter tip to generate photoelectron pulses accelerated to 30 keV. With the BP specimens mounted on a high precision manipulator with five axes of motion, suitable sample regions for stable stroboscopic measurements were found by fine translational adjustments; different orders of out-of-plane (0*k*0) diffractions were obtained by changing the incidence angle relative to the fixed electron beam. In the present study, the UED measurements were conducted at a repetition rate of 10 kHz. Additionally, a pulse-front tilt setup has been implemented to overcome the temporal mismatch between the arrivals of the near-vertically descending excitation and grazing electron beams (*37, 38*), which led to the improved instrumental response time of ~500 fs (Fig. S6). Moreover, the matching of the laser and photoelectron arrival times has the benefit of a well-defined zero of time (within an error range up to 100 fs in a poorly-overlapped scenario; see Supplementary Materials and Fig. S7), which is crucial to confident identification of any onset delays in the dynamics. The laser fluences were estimated using the full width at half maximum of the laser footprint of 200×1500 μm² on the specimen, which is sufficiently larger to cover the tightly focused electron footprint of 15×230 μm² for a uniform photoexcited region. The resulting diffraction images at different delay times were formed on a phosphorus screen and captured by an intensified CMOS camera.

Further details about the calculations of photoinjected carrier densities, lattice temperature increases, Debye–Waller analysis, and thermal diffusion can be found in the Supplementary Materials.

**Acknowledgments**

The authors wish to thank N. Z. Wang and X. H. Chen of USTC for providing BP samples and F. Wang for initial optical measurements. This research was primarily supported by the R. A. Welch Foundation (E-1860). X.H. and the instrumental implementation of the pulse-front tilt scheme were partly supported by a National Science Foundation CAREER Award (CHE-1653903). The initial support by the Samsung Advanced Institute of Technology's Global Research Outreach (GRO) Program is acknowledged.